\documentclass[12pt,a4paper]{article}

\usepackage{ifthen} 
\newboolean{pdflatex}
\setboolean{pdflatex}{true} 

\newboolean{articletitles}
\setboolean{articletitles}{true} 

\newboolean{uprightparticles}
\setboolean{uprightparticles}{false} 

\newboolean{inbibliography}
\setboolean{inbibliography}{false} 
\usepackage{multirow}

\newcommand{\pHe}{\ensuremath{p\text{He}}\xspace}
\newcommand{\pAr}{\ensuremath{p\text{Ar}}\xspace}

\def\paperauthors{LHCb collaboration} 
\def\paperasciititle{First measurement of charm production in fixed-target configuration at the LHC} 
\def\papertitle{First measurement of charm production in fixed-target configuration at the LHC} 
\def\paperkeywords{{High Energy Physics}, {LHCb}} 
\def\papercopyright{2018 CERN for the benefit of the LHCb collaboration}
\def\paperlicence{CC-BY-4.0}
\def\paperlicenceurl{https://creativecommons.org/licenses/by/4.0/}

\usepackage[top=1in, bottom=1.25in, left=1in, right=1in]{geometry}
\usepackage{microtype}
\usepackage{lineno}  
\usepackage{xspace} 
\usepackage{caption} 

\usepackage{graphicx}  
\usepackage{color}
\usepackage{colortbl}
\graphicspath{{./figs/}} 

\usepackage{amsmath} 
\usepackage{amssymb}
\usepackage{amsfonts}
\usepackage{upgreek} 

\newcommand*\patchAmsMathEnvironmentForLineno[1]{%
\expandafter\let\csname old#1\expandafter\endcsname\csname #1\endcsname
\expandafter\let\csname oldend#1\expandafter\endcsname\csname
end#1\endcsname
 \renewenvironment{#1}%
   {\linenomath\csname old#1\endcsname}%
   {\csname oldend#1\endcsname\endlinenomath}%
}
\newcommand*\patchBothAmsMathEnvironmentsForLineno[1]{%
  \patchAmsMathEnvironmentForLineno{#1}%
  \patchAmsMathEnvironmentForLineno{#1*}%
}
\AtBeginDocument{%
\patchBothAmsMathEnvironmentsForLineno{equation}%
\patchBothAmsMathEnvironmentsForLineno{align}%
\patchBothAmsMathEnvironmentsForLineno{flalign}%
\patchBothAmsMathEnvironmentsForLineno{alignat}%
\patchBothAmsMathEnvironmentsForLineno{gather}%
\patchBothAmsMathEnvironmentsForLineno{multline}%
\patchBothAmsMathEnvironmentsForLineno{eqnarray}%
}

\usepackage{hyperxmp}

\usepackage[pdftex,
            pdfauthor={\paperauthors},
            pdftitle={\paperasciititle},
            pdfkeywords={\paperkeywords},
            pdfcopyright={Copyright (C) \papercopyright},
            pdflicenseurl={\paperlicenceurl}]{hyperref}

\usepackage[all]{hypcap} 

\def\lhcb   {\mbox{LHCb}\xspace}
\def\pythia     {\mbox{\textsc{Pythia}}\xspace}
\def\photos     {\mbox{\textsc{Photos}}\xspace}
\def\geant      {\mbox{\textsc{Geant4}}\xspace}
\def\sqsnn {\ensuremath{\protect\sqrt{s_{\scriptscriptstyle\rm NN}}}\xspace}
\def\sqs   {\ensuremath{\protect\sqrt{s}}\xspace}
\newcommand{\stat}{\ensuremath{\mathrm{\,(stat)}}\xspace}
\newcommand{\syst}{\ensuremath{\mathrm{\,(syst)}}\xspace}
\newcommand{\tev}{\ifthenelse{\boolean{inbibliography}}{\ensuremath{~T\kern -0.05em eV}}{\ensuremath{\mathrm{\,Te\kern -0.1em V}}}\xspace}
\newcommand{\gev}{\ensuremath{\mathrm{\,Ge\kern -0.1em V}}\xspace}
\newcommand{\gevc}{\ensuremath{{\mathrm{\,Ge\kern -0.1em V\!/}c}}\xspace}
\newcommand{\gevcc}{\ensuremath{{\mathrm{\,Ge\kern -0.1em V\!/}c^2}}\xspace}
\newcommand{\mev}{\ensuremath{\mathrm{\,Me\kern -0.1em V}}\xspace}
\newcommand{\mevc}{\ensuremath{{\mathrm{\,Me\kern -0.1em V\!/}c}}\xspace}
\newcommand{\mevcc}{\ensuremath{{\mathrm{\,Me\kern -0.1em V\!/}c^2}}\xspace}
\def\nb {\ensuremath{\mathrm{ \,nb}}\xspace}
\def\mm {\ensuremath{\mathrm{ \,mm}}\xspace}
\def\pt {\ensuremath{p_{\mathrm{T}}}\xspace}
\def\ps {\ensuremath{{\mathrm{ \,ps}}}\xspace}
\def\mub{\ensuremath{{\mathrm{ \,\upmu b}}}\xspace}
\def\PJ        {\ensuremath{J}\xspace} 
\def\Ppsi      {\ensuremath{\psi}\xspace}                 
\def\jpsi      {{\ensuremath{{\PJ\mskip -3mu/\mskip -2mu\Ppsi\mskip 2mu}}}\xspace}
\def\PD        {\ensuremath{D}\xspace}                 
\def\D         {{\ensuremath{\PD}}\xspace}
\def\Dz        {{\ensuremath{\D^0}}\xspace}
\def\Dbar    {{\kern 0.2em\overline{\kern -0.2em \PD}{}}\xspace}

\def\Km      {{\ensuremath{\kaon^-}}\xspace}
\def\pip    {{\ensuremath{\pion^+}}\xspace}
\def\PK      {\ensuremath{K}\xspace}                 
\def\kaon    {{\ensuremath{\PK}}\xspace}
\def\Ppi         {\ensuremath{\pi}\xspace}                 
\def\pion   {{\ensuremath{\Ppi}}\xspace}

\def\Pc        {\ensuremath{c}\xspace}                 
\def\cquark    {{\ensuremath{\Pc}}\xspace}
\def\cquarkbar {{\ensuremath{\overline \cquark}}\xspace}
\def\ccbar     {{\ensuremath{\cquark\cquarkbar}}\xspace}
\def\Pb        {\ensuremath{b}\xspace}                 
\def\bquark    {{\ensuremath{\Pb}}\xspace}
\def\Pp        {\ensuremath{p}\xspace}                 
\def\proton    {{\ensuremath{\Pp}}\xspace}
\def\Pmu         {\ensuremath{\mu}\xspace}                 
\def\mup        {{\ensuremath{\Pmu^+}}\xspace}
\def\mun        {{\ensuremath{\Pmu^-}}\xspace} 

\usepackage{cite} 
\usepackage{mciteplus}

\begin{document}

\renewcommand{\thefootnote}{\fnsymbol{footnote}}
\setcounter{footnote}{1}

\begin{titlepage}
\pagenumbering{roman}

\vspace*{-1.5cm}
\centerline{\large EUROPEAN ORGANIZATION FOR NUCLEAR RESEARCH (CERN)}
\vspace*{1.5cm}
\noindent
\begin{tabular*}{\linewidth}{lc@{\extracolsep{\fill}}r@{\extracolsep{0pt}}}
\ifthenelse{\boolean{pdflatex}}
 & & CERN-EP-2018-266 \\  
 & & LHCb-PAPER-2018-023 \\  
 & & \today \\ 
 & & \\
\end{tabular*}

\vspace*{4.0cm}

{\normalfont\bfseries\boldmath\huge
\begin{center}
  \papertitle 
\end{center}
}

\vspace*{2.0cm}

\begin{center}
\paperauthors\footnote{Full author list given at the end of the Letter.}
\end{center}

\vspace{\fill}

\begin{abstract}
  \noindent
The first measurement of heavy-flavor production by the \lhcb experiment in its fixed-target mode is presented.
The production of \jpsi and \Dz\ mesons is studied with beams of protons of different energies colliding with gaseous targets of helium and argon with nucleon-nucleon center-of-mass energies of $\sqsnn = 86.6 $ and $ 110.4 \gev$, respectively. The \jpsi\ and \Dz\ (including charge conjugate) production cross sections in $p{\rm He}$ collisions in the rapidity range $[2,4.6]$ are found to be $\sigma_{\jpsi} = 652 \pm 33\stat\pm 42\syst$ nb$/$nucleon and $\sigma_{D^0} = 80.8 \pm 2.4 \stat \pm 6.3 \syst$  $\mu$b$/$nucleon, where the first uncertainty is statistical and the second is systematic. No evidence for a substantial intrinsic charm content of the nucleon is observed in the large Bjorken-$x$ region. 
\end{abstract}

\vspace*{2.0cm}
\begin{center}
Published in Phys.~Rev.~Lett. \textbf{122}, 132002 (2019)
\end{center}

\vspace{\fill}
{\footnotesize 
\centerline{\copyright~\papercopyright. \href{\paperlicenceurl}{\paperlicence} licence.}}
\vspace*{2mm}
\end{titlepage}
\newpage
\setcounter{page}{2}
\mbox{~}
\cleardoublepage

\renewcommand{\thefootnote}{\arabic{footnote}}
\setcounter{footnote}{0}


\pagestyle{plain} 
\setcounter{page}{1}
\pagenumbering{arabic}
In the high-density and high-temperature regime of quantum chromodynamics (QCD), the production of heavy quarks  in nucleus-nucleus interactions is well suited to study the transition between ordinary hadronic matter and the hot and dense quark-gluon plasma (QGP). 
Heavy quarks are produced only in the early stages of the interaction, because their masses are significantly higher than the QGP critical temperature, $T_c\sim156 \mev$\cite{Steinbrecher}. 
Lattice QCD predictions imply that, at sufficiently high temperature, the production of heavy quark-antiquark bound states decreases due to the modification of their binding mechanism\cite{Digal2005}. 

The interpretation of the charmonium \ccbar bound states suppression, observed in nucleus-nucleus collisions at various energies\cite{Kluberg}, can be significantly sharpened by measuring charmonium yields together with the overall charm quark production\cite{Satz_1994}.  
 The production of \Dz mesons, made of a $c$ and a $\bar{u}$ quark, reflects a large fraction of the overall charm quark production. 
The study of charmonium production in proton-nucleus collisions on various nuclear targets, where no QGP is formed, is needed to establish the charmonium suppression patterns observed in heavy-ion collisions and to understand the mechanisms underlying charmonium production~\cite{ConesadelValle:2011fw, Ferreiro2009}. 

Several effects can be studied in proton-nucleus collisions, such as the interaction of \ccbar pairs with the target nucleons leading to a breakup of the charmonium states~\cite{Lourenco}, parton shadowing (or antishadowing) in the target nucleus~\cite{deFlorian:2003qf, Eskola:2009uj} that modifies charmonium production, saturation effects~\cite{Alba2014}, and parton energy loss~\cite{Arleo_Peigne, Arleo2013, Arleo_Peigne_Sami}. 

In this Letter, the first measurement of heavy-flavor production in a fixed-target mode at the LHC is presented. The production of \jpsi and \Dz mesons  are studied in collisions of protons with energies of 4 \tev\ and 6.5 \tev\ incident on helium and argon nuclei at rest at center-of-mass energies of $\sqsnn~=~86.6~\gev$ and $\sqsnn~=~110.4~\gev$, respectively.

The \lhcb detector~\cite{Alves:2008zz,LHCb-DP-2014-002} is a single-arm forward spectrometer covering the pseudorapidity range $2 < \eta < 5$, designed for the study of particles containing \cquark\ or \bquark\ quarks. The detector elements that are particularly relevant to this analysis are: the VELO surrounding the \proton\proton interaction region that allows \cquark\ and \bquark\ hadrons to be identified from their characteristic flight distance; a tracking system that provides a measurement of the momentum of charged particles; two ring-imaging Cherenkov detectors that are able to discriminate between different species of charged hadrons; a calorimeter system consisting of scintillating-pad and preshower detectors, electromagnetic and hadronic calorimeters; and a muon detector composed of alternating layers of iron and multiwire proportional chambers.
The system for measuring overlap with gas (SMOG) device~\cite{LHCb-PAPER-2014-047} enables the injection of gases with pressure of $\mathcal{O}(10^{- 7})$ mbar in the beam pipe section crossing the  silicon-strip vertex locator (VELO), allowing LHCb to operate as a fixed-target experiment.
SMOG allows the injection of noble gases and therefore gives the unique opportunity to study nucleus-nucleus and proton-nucleus collisions on various targets. 
Thanks to the boost induced by the high-energy proton beam the \lhcb acceptance covers the backward rapidity hemisphere in the center-of-mass system of the reaction from a very negative center-of-mass rapidity $y^*\sim-2.5$ to $y^*\sim0$. Therefore, the SMOG fixed-target program offers many new opportunities of physics studies~\cite{Brodsky13}, including the study of heavy-quark production in the large Bjorken-$x$ region, with $x$ the fraction of the nucleon momentum carried by the target parton, up to $\sim0.37$ for \Dz mesons, and the test of the intrinsic charm content of the proton~\cite{Pumplin07,Dulat14}. 

The events are triggered by the two-stage trigger system of the experiment~\cite{trigger}.
The first level is implemented in hardware and uses information provided by the calorimeters and the muon detectors, while the second is a software trigger. 
The hardware trigger requires at least one identified muon for the selection of the $\jpsi\to\mun\mup$ candidates, and  a minimal activity in the calorimeter for the $\Dz\to K^- \pi^+$ selection (inclusion of charge conjugate processes is implied throughout).  The software trigger requires two well-reconstructed muons forming an invariant mass larger than 2700\mevcc for the $\jpsi$ selection. For the \Dz selection, it requires a well-reconstructed vertex formed by well-identified kaon and pion tracks, both of which are required to have a transverse momentum larger than 500\mevc and an invariant mass between 1715 and 2015\mevcc.

The data samples have been collected under particular beam conditions where proton bunches moving towards the detector do not cross any bunch moving in the opposite direction at the nominal $pp$ interaction point. 
Events with \jpsi or \Dz candidates must have a reconstructed primary vertex within the fiducial region $-200\mm<z_{PV}<200\mm$, where high reconstruction efficiencies are achieved and calibration samples available.
In order to suppress residual $pp$ collisions, events with activity in the backward region are vetoed, based on the number of hits in VELO stations upstream of the interaction region. 

The offline selection of \jpsi and \Dz candidates is similar to that used in Refs.~\cite{Aaij:2013zxa, LHCb-CONF-2016-003}. Specifically, events with at least one primary vertex are selected where the primary vertex is reconstructed from at least four tracks in the VELO detector. The \jpsi candidates are obtained from two oppositely signed muons forming a good-quality vertex. The well-identified muons have a transverse momentum, \pt, larger than 700 \mevc and are required to be consistent  with originating from the primary interaction point.  The kaon and pion  from the \Dz decay are required to be of good quality and to come from a common displaced vertex.  Tight requirements are set on the kaon and pion particle identification criteria. The \Dz\ candidates are selected to have a decay time larger than 0.5\ps. The measurements are performed in the range of \jpsi and \Dz\ transverse momentum \pt~$<~8\gevc$ and rapidity $2.0 < y < 4.6$. 

Acceptance and reconstruction efficiencies are determined using simulated $p$He and $p$Ar events.
                                                          
In the simulation, \jpsi and \Dz mesons are generated using \pythia~8~\cite{Pythia8, Sjostrand:2006za}  with a specific \lhcb configuration~\cite{1742-6596-331-3-032047} and with colliding-proton beam momenta being equal to the momenta per nucleon of the beam and target in the center-of-mass frame. 
The decays  are described by {\sc EvtGen}~\cite{EvtGen}, in which  final-state  radiation is generated using \photos~\cite{PHOTOS}. The four-momenta of the \jpsi and \Dz\  daughters are then extracted and embedded into \pAr\ or \pHe\ minimum bias events that are generated with the EPOS event generator~\cite{Pierog:2013ria} using beam parameters obtained from the data.   Decays of hadronic particles generated with EPOS are also described by {\sc EvtGen}. The  interaction of the generated particles with the detector, and its response, are implemented using the \geant toolkit~\cite{geant2, geant4} as described in Ref.~\cite{gauss}.

The \jpsi detection efficiency is dependent on its polarization. Since no polarization measurement has yet been made for data collected at \sqsnn close to 100 \gev, the polarization is assumed to be zero and no corresponding systematic uncertainty is quoted on the cross-section results. 
A small longitudinal polarization, described by the parameter $\lambda_\theta$, has been found at different energies close to $\lambda_\theta=-0.1$~\cite{LHCb-PAPER-2013-008, PHENIXpol, HERABpol}. Using data from Ref.\cite{LHCb-PAPER-2015-037} and assuming a value $\lambda_\theta = -0.1$, the measured \jpsi cross section would decrease by about 1$\%$ to 2.3$\%$ depending on the \jpsi ($\pt$, $y$) bin\cite{supplemental}.

The prompt \jpsi\ and \Dz\ signal yields are obtained from extended unbinned maximum-likelihood fits to their mass distributions. 
\begin{figure}[tb]
\begin{center}
\includegraphics[width=0.49\textwidth]{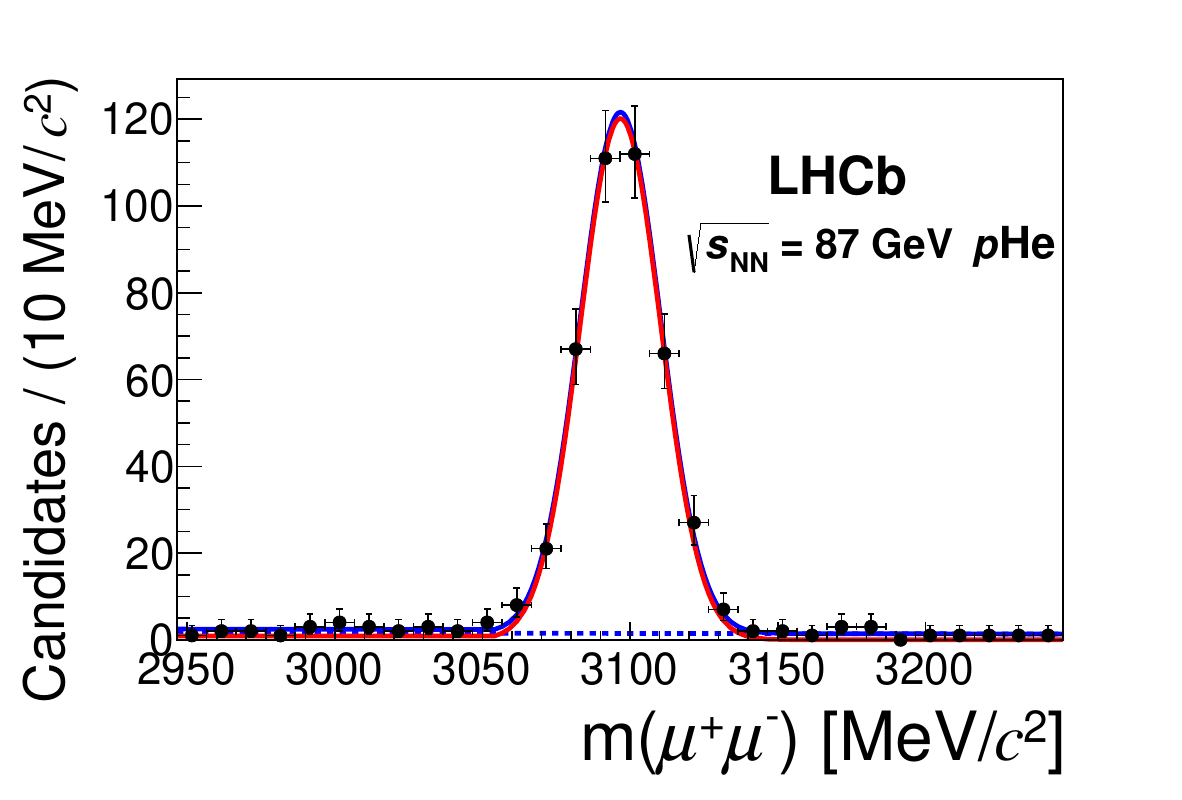}
\includegraphics[width=0.49\textwidth]{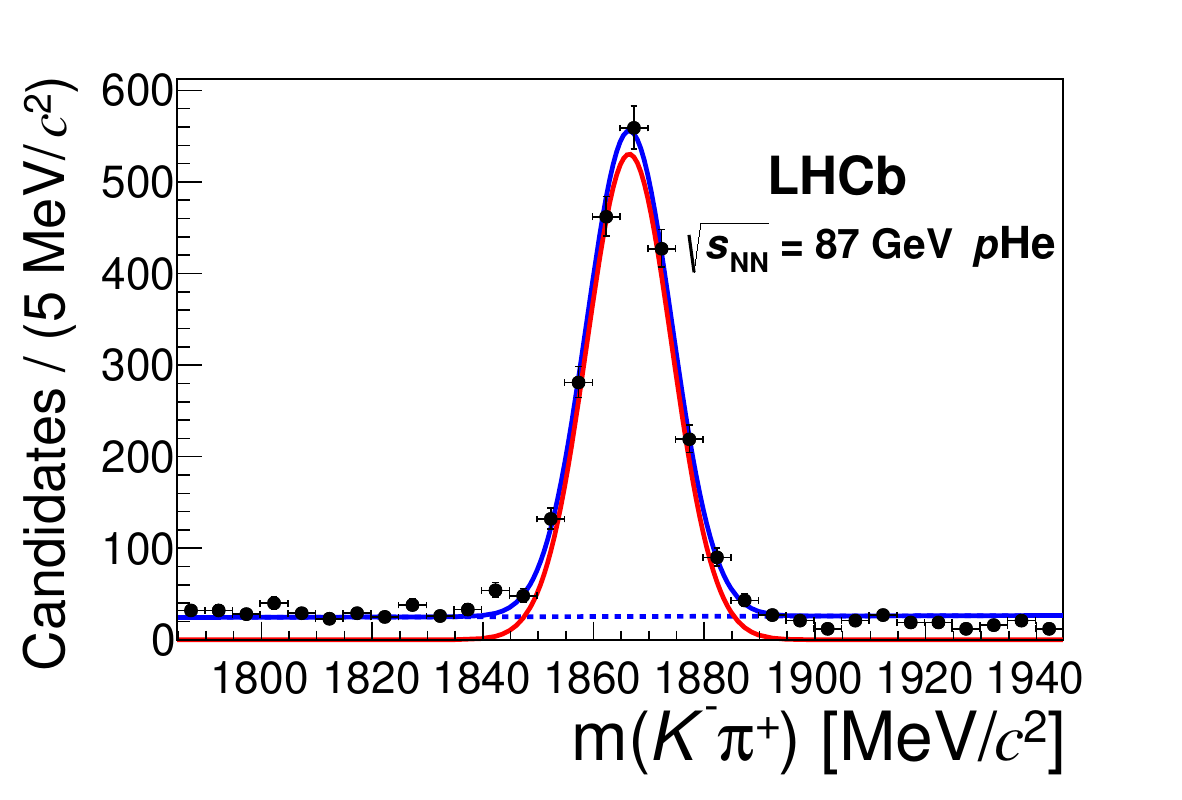}
\caption{Mass distributions, fitted by an extended unbinned maximum-likelihood in \mbox{$\sqsnn=86.6\gev$} \pHe collisions;  $\jpsi\to\mun\mup$ (left); $\Dz\to\Km\pip$ (right). The dashed blue line corresponds to the combinatorial background, the red line to the signal and the solid blue line to the sum of the two.}
\label{fig::sig_pHe}
\end{center}
\end{figure}
The fit functions are given by the sum of a crystal ball function~\cite{CB} describing the \jpsi signal, and an exponential function for the background. The \Dz signal is fitted by the sum of two Gaussian functions, and an exponential function for the background. Figure~\ref{fig::sig_pHe} shows the mass distributions obtained after all selection criteria are applied to the entire \pHe\ data set, with the fit functions superimposed.

The signal yields are determined in uniformly populated bins of \pt\ or $y$. 
A coarser binning scheme is used for \jpsi\ candidates, owing  to the smaller sample size.
The yields determined from the mass fit are corrected for the total efficiencies, which include the geometrical acceptance of the detector, the event trigger, the event selection, the primary vertex, the track reconstruction, and particle identification. Particle identification~\cite{LHCb-PUB-2016-021}  and tracking efficiencies are obtained 
from control sample of \proton\proton collision data. All the other efficiencies are determined from simulation.
Several sources of systematic uncertainties are considered, affecting either the determination of the signal yields or the total efficiencies. They are summarised in Table~\ref{tab:systematics} separately for correlated and uncorrelated systematic uncertainties.
\begin{table}[t]
\caption{\small \label{tab:systematics} Systematic and statistical uncertainties on the \jpsi and \Dz yields in $\%$. Systematic uncertainties correlated between bins affect all measurements by the same relative amount. Ranges denote the minimum and the maximum values among the $y$ or \pt bins.}
\begin{center}
{\small
\begin{tabular}{@{}llll@{}}
\hline
Source & & \jpsi & \Dz\  \\  
\hline
{\textbf{Correlated between bins}}\span\omit & &  \\
\multirow{2}*{Signal selection efficiency}  & \pAr & 1.4$\%$ & 1.4$\%$ \\
   & \pHe & 1.1$\%$ & 1.1$\%$  \vspace{0.1cm} \\
\multirow{2}*{Tracking efficiency} & \pAr            & 1.9$\%$ & 3.5$\%$ \\
                        & \pHe            & 1.1$\%$  & 3.2$\%$  \vspace{0.1cm} \\
Particle identification  & \pAr                   & (1.8 -- 1.9)$\%$  & (4.3 -- 5.4)$\%$ \\
efficiency                   & \pHe                   & (0.9 -- 1.0)$\%$ & (1.1 -- 2.6) $\%$  \vspace{0.1cm} \\
\hline
\hline
{\textbf{Uncorrelated between bins}}\span\omit & & \\
\multirow{2}*{Signal determination} & \pAr & (0 -- 0.9)$\%$ & (1.6 -- 2.6)$\%$  \\
& \pHe & (0 -- 0.9)$\%$ & (1.6 -- 2.5)$\%$ \vspace{0.1cm} \\
\multirow{2}*{Tracking efficiency} & \pAr   & (0.1 -- 1.9)$\%$ & (0.2 -- 2.6)$\%$  \\
                        & \pHe            & (0.2 -- 1.8)$\%$ & (0.3 -- 2.7)$\%$  \vspace{0.1cm} \\
\multirow{2}*{Simulation sample} & \pAr & (1.8 -- 2.0)$\%$ & (2.4 -- 2.5)$\%$ \\ 
                         & \pHe & (1.7 -- 3.4)$\%$ & (2.5 -- 2.8)$\%$  \vspace{0.1cm} \\ 
Particle identification & \pAr                   & (0 -- 1.9)$\%$ & (0 -- 5.6)$\%$ \\
efficiency                   & \pHe                   & (0 -- 0.8)$\%$ & (0 -- 3.7)$\%$  \vspace{0.1cm} \\
\hline
\multirow{2}*{Statistical uncertainties} & \pAr & (7.8 -- 12.7)$\%$ & (2.8 -- 5.8)$\%$ \\
                                         & \pHe & (7.9 -- 11.3)$\%$ & (4.2 -- 10.1)$\%$  \\
\hline
\end{tabular} 
}
\end{center}
\end{table}

A systematic uncertainty is assigned to the signal determination. A first contribution, common to \jpsi and \Dz signals, is obtained by determining the maximum contamination from residual \proton\proton collisions. The systematic uncertainty related to the determination of the signal yields includes the contribution from $b$-hadron decays and the mass fit. The fraction of signal from \bquark hadrons, determined through the fit of the impact parameter distribution of the \Dz candidates with respect to the primary vertex, is $(0.9^{+1.6}_{-0.9}) \%$. The systematic uncertainty related to the mass fit is evaluated using alternative models for signal and background shapes that reproduce the mass shapes equally well. 

Another source of uncertainty is associated with the accuracy of the simulation used to compute the acceptances and efficiencies. This systematic uncertainty includes the statistical uncertainty due to the finite size of the simulation sample and the differences in the distributions of the transverse momentum and rapidity between data and simulation. This systematic uncertainty is computed in each $y$ and \pt bin.
Systematic uncertainties in tracking and particle identification efficiencies are mainly related to the differences between the track multiplicity in \pAr, \pHe\ and \proton\proton collisions. The tracking systematic uncertainty also takes into account the difference in tracking efficiency between the data and the simulation.

\begin{figure}[t]
\begin{center}
\includegraphics[width=0.49\textwidth]{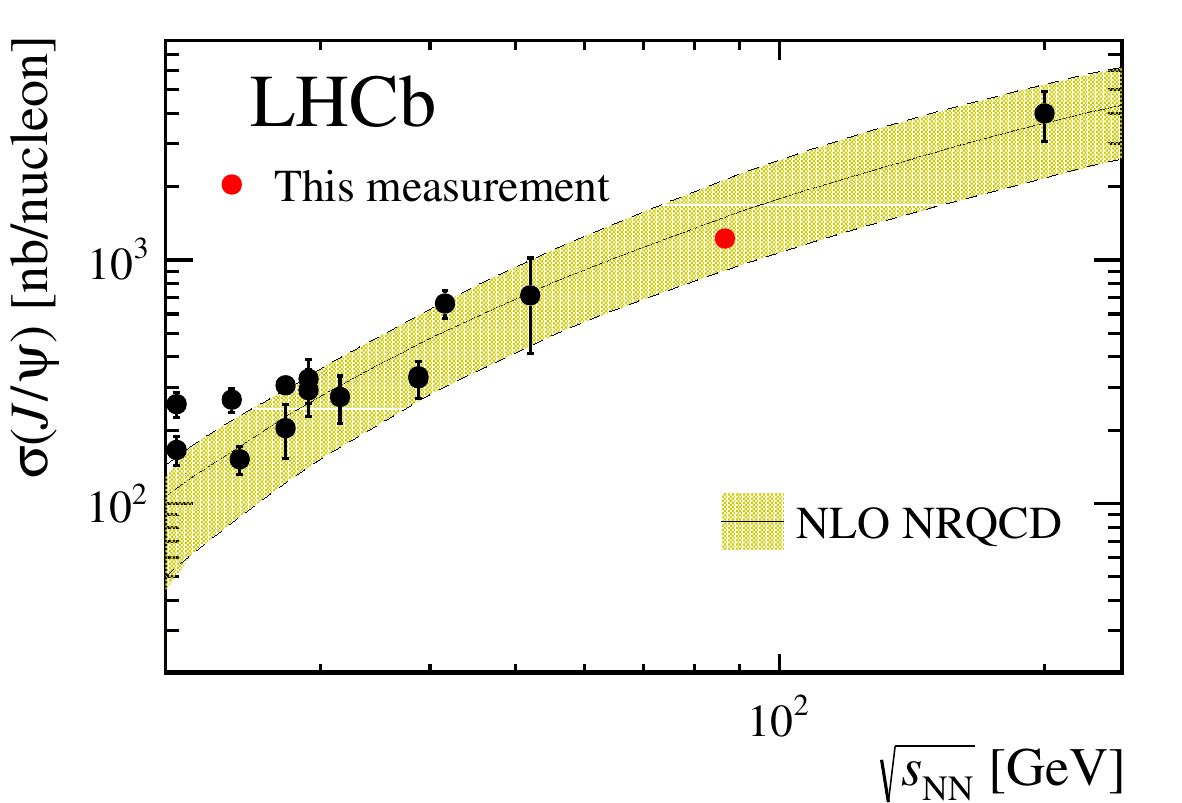}
\includegraphics[width=0.49\textwidth]{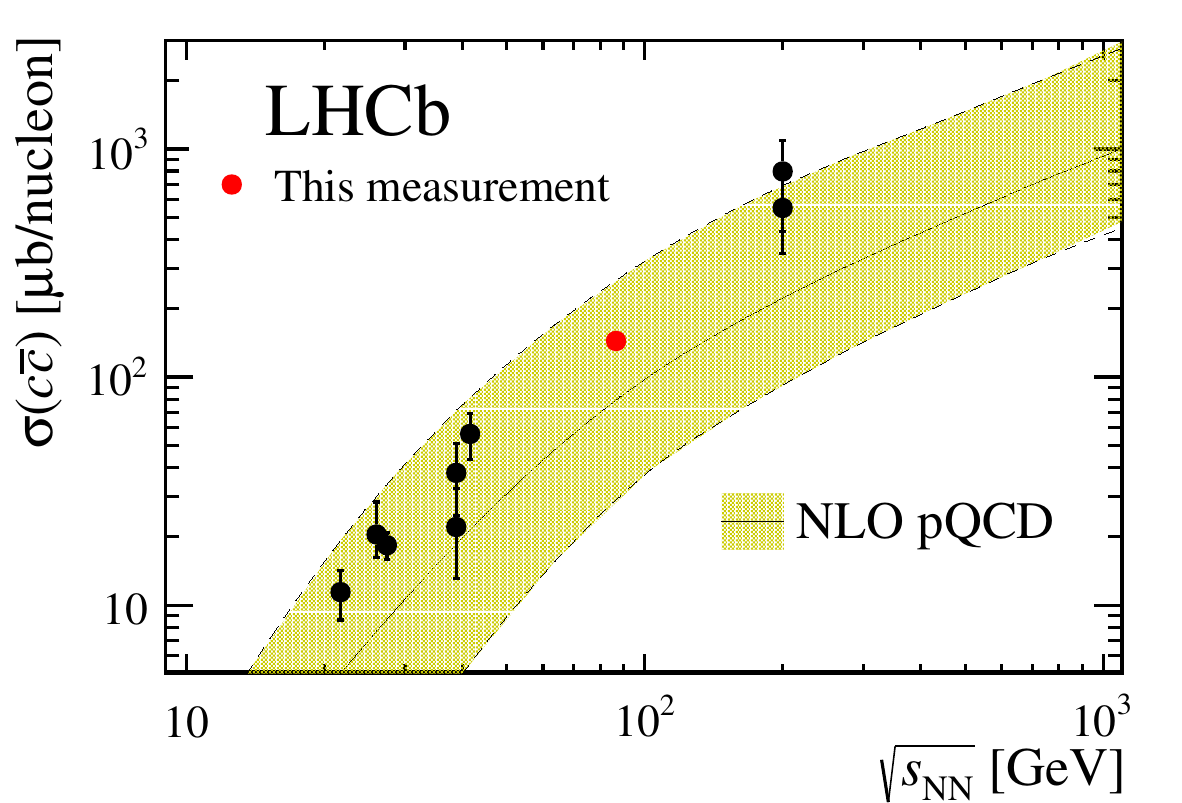}
\caption{Left: \jpsi\ cross-section measurements as a function of the center-of-mass energy. Experimental data, represented by black points, are taken from Ref.~\cite{Maltoni:2006yp}. The band corresponds to a fit based on NLO NRQCD calculations~\cite{Maltoni:2006yp}. Right: $c\overline{c}$ cross-section measurements as a function of the center-of-mass energy. Experimental data, represented by black points, are taken from Ref.~\cite{Adam:2016ich}. The yellow band corresponds to NLO pQCD calculations~\cite{Mangano:92}. Red points correspond to the \pHe\ results from the present analysis.}
\label{fig::cross-section}
\end{center}
\end{figure}
The cross-section measurement is made for the $p$He sample only, for which the luminosity determination is available.
The luminosity is determined from the yield of electrons elastically scattering off the target He atoms~\cite{antiproton} to be ${\cal{L}}_{p\textrm{He}} = 7.58 \pm 0.47$ nb$^{-1}$. The measured \jpsi\ and \Dz\ cross sections per target nucleon within $y\in [2,4.6]$, after correction for the branching fractions $\jpsi \to \mu^+ \mu^-$ and $\Dz \to K^- \pi^+$ (and charge conjugate), are
\begin{equation*}
\begin{split}
&\sigma_{\jpsi} = 652 \pm 33 \stat  \pm 42 \syst\nb/\mathrm{nucleon},\\
&\sigma_{D^0} = 80.8 \pm 2.4 \stat \pm 6.3 \syst\mub/\mathrm{nucleon}.
\end{split}
\end{equation*}
In order to compare to  previous experimental results at different energies, both  \jpsi and \Dz cross sections are extrapolated to the full phase-space using \pythia 8 with a specific LHCb tuning and with the CT09MCS PDF set~\cite{CT09MCS}. The extrapolation factor is $2f$, where $f=0.940$ for the \jpsi and $f=0.965$ for the \Dz, and describes the extrapolation from $y^*\in [-2.53, 0.07]$ to the full backward (negative) rapidity hemisphere, assuming forward-backward symmetry. The full phase-space cross sections are
\begin{equation*}
\begin{split}
&\sigma_{\jpsi} = 1225.6 \pm 100.7\nb/\mathrm{nucleon},\\
&\sigma_{D^0} = 156.0 \pm 13.1\mub/\mathrm{nucleon},
\end{split}
\end{equation*}
where statistical and systematic uncertainties have been added quadratically and no systematic uncertainties due to the extrapolation are included. 
In addition, the \Dz cross section is scaled with the global fragmentation factor $f(c \to D^0) = 0.542 \pm 0.024$~\cite{Gladilin:2014tba}, in order to obtain the $c\overline{c}$ production cross section: 
\begin{equation*}
    \sigma_{c\overline{c}} = \frac{\sigma_{\Dz}}{2\times f(c \to \Dz)}= 144 \pm 12 \pm 4\mub/\mathrm{nucleon}. 
\end{equation*}
The last uncertainty reflects the limited knowledge of the fragmentation factor.
An overview of \jpsi and $c\overline{c}$ cross-section measurements at different center-of-mass energies by different experiments are shown in Fig.~\ref{fig::cross-section} including this measurement. 
The \jpsi\ cross section is compared to a fit based on NLO NRQCD calculations~\cite{Maltoni:2006yp} and the $c\overline{c}$ cross section to NLO pQCD calculations~\cite{Adam:2016ich,Mangano:92}. The \ccbar cross section shows a small tension with respect to theoretical calculations as already observed at 200 \gev, while the \jpsi\ cross-section measurement is in good agreement with the fit based on NLO pQCD calculations.
 \begin{figure}[t]
 \begin{center}
 \includegraphics[width=0.49\textwidth]{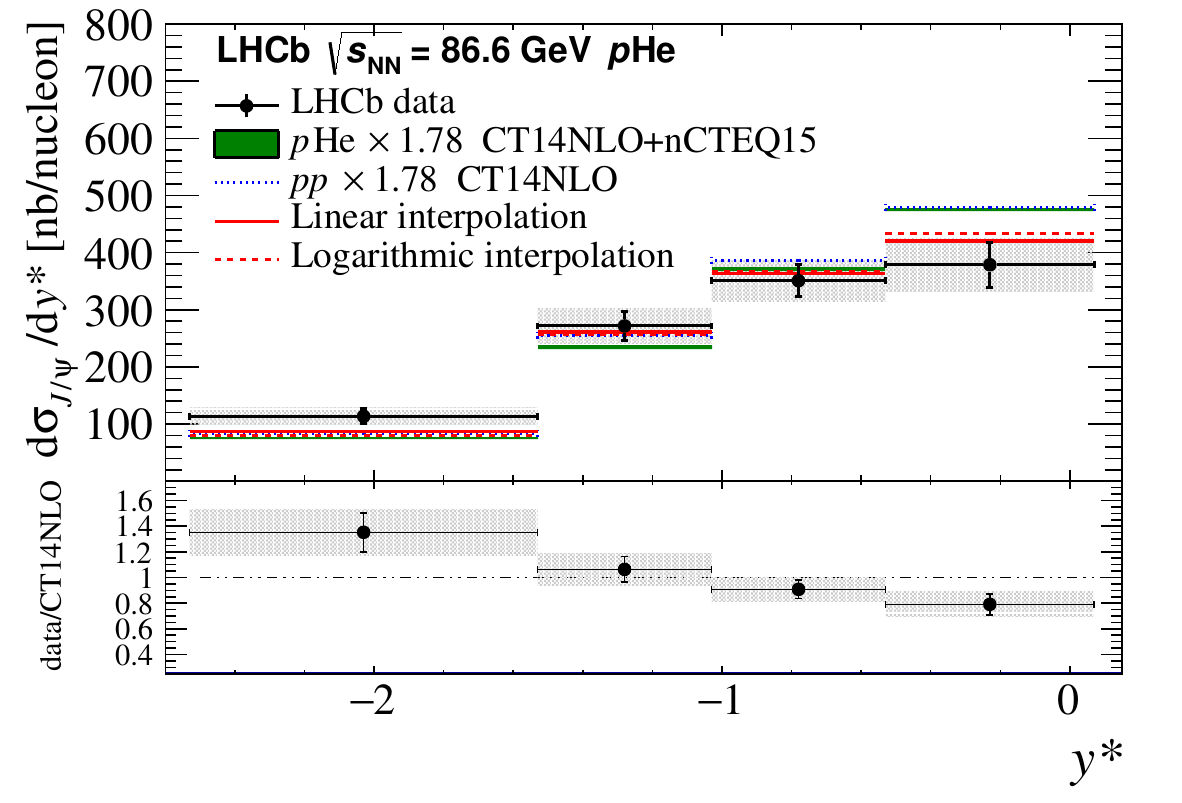}
 \includegraphics[width=0.49\textwidth]{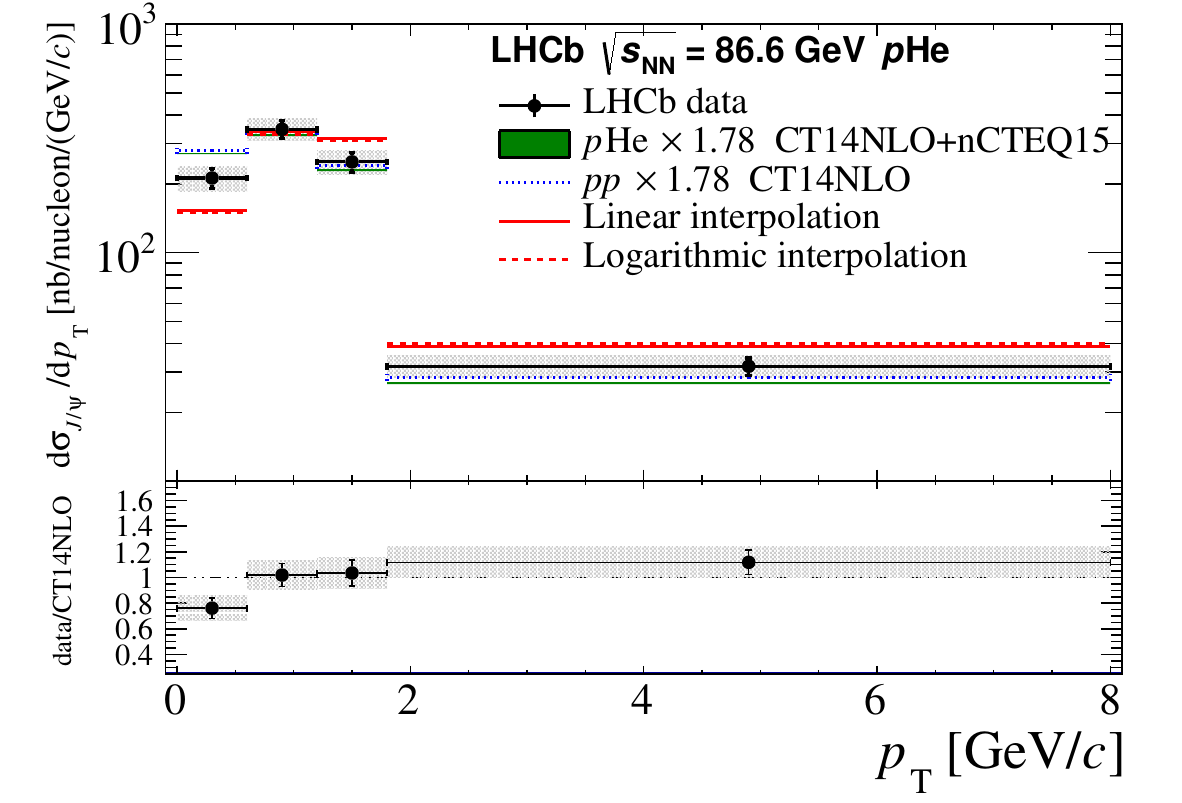}
 \includegraphics[width=0.49\textwidth]{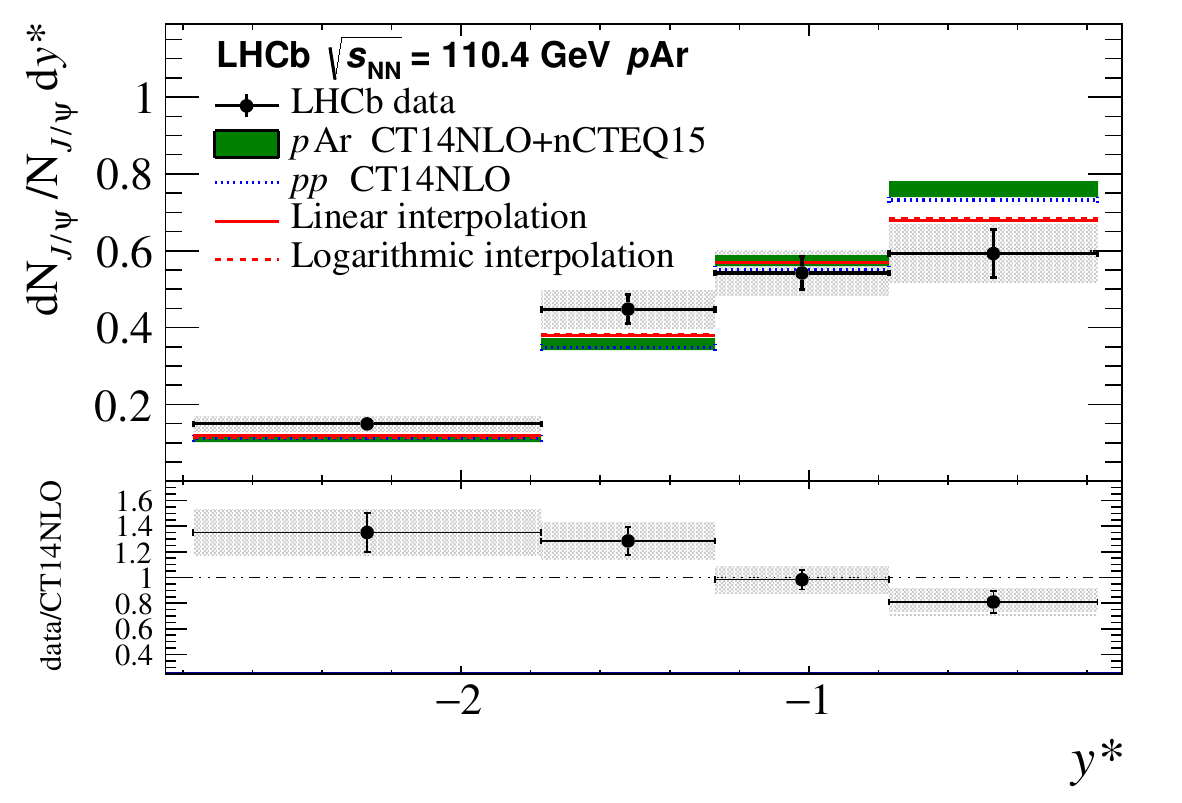}
 \includegraphics[width=0.49\textwidth]{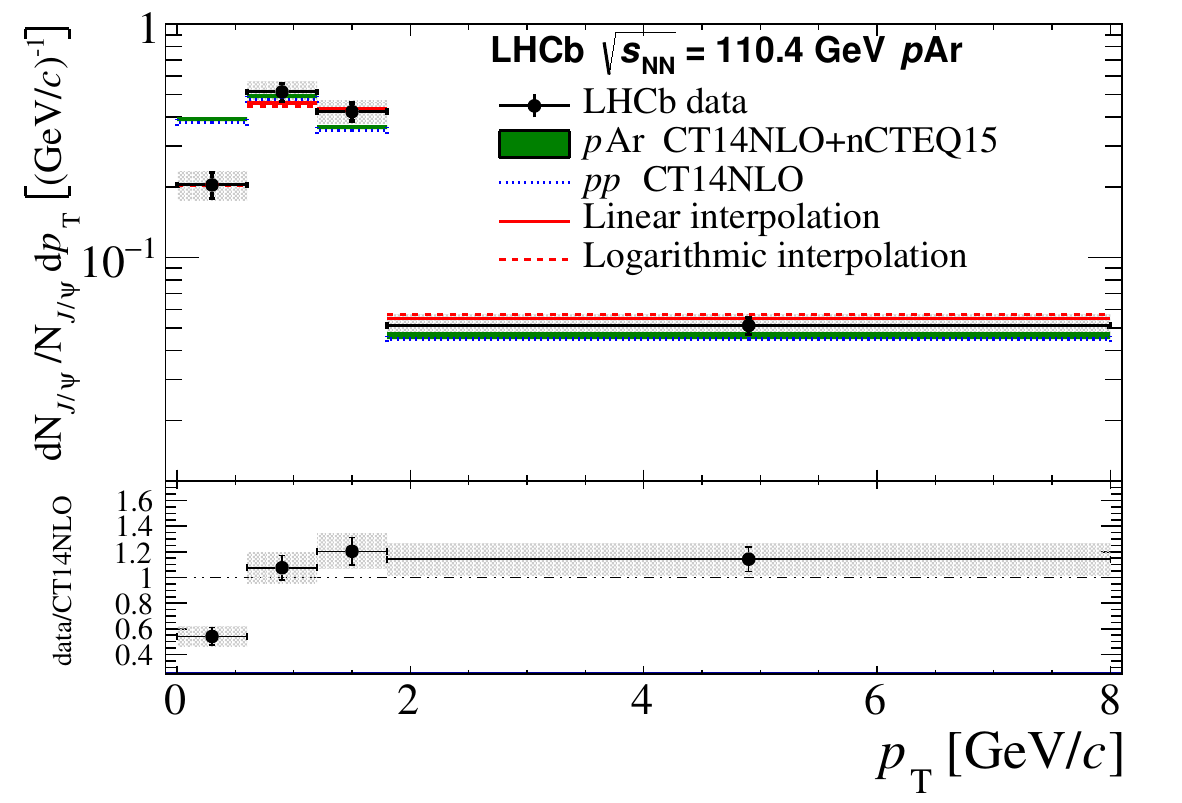}
 \caption{Differential \jpsi\ production cross sections for (top) \pHe\ and differential \jpsi\ yields for (bottom) \pAr\ collisions, as a function of (left) center-of-mass rapidity $y^*$ and (right) transverse momentum $\pt$. The data points mark  the bin centers. The quadratic sum  of statistical and uncorrelated systematic uncertainties are indicated by the vertical black lines. The correlated systematic uncertainties are indicated by the grey area. Theoretical predictions are described in the text. The lower panel of each plot shows the ratio of data to HELAC-ONIA $pp$ predictions.}
 \label{fig::sig_jpsi}
 \end{center}
 \end{figure}
The \jpsi\ differential cross sections per target nucleon obtained for the \pHe\ dataset, as functions of $y^*$ and $\pt$, are shown in the two top plots of Fig.~\ref{fig::sig_jpsi} and given in Ref.\cite{supplemental}. These results are compared with HELAC-ONIA predictions~\cite{Lansberg:2016deg,Shao:2015vga,Shao:2012iz}, for $pp$ (CT14NLO PDF set~\cite{Dulat16}) and \pHe\ (CT14NLO+nCTEQ15 PDF~\cite{nCTEQ15} sets) collisions. The predictions underestimate the measured total cross section. The HELAC-ONIA predictions are rescaled by a factor 1.78 in Fig.~\ref{fig::sig_jpsi} to compare  the shape of the distributions. Data are also compared with  phenomenological parametrizations, interpolated to the present data energies, based on Refs.~\cite{ArleoPeigne1303,Arleo2013}. Solid and dashed red lines are obtained with linear and logarithmic interpolations, respectively, between the results from the E789 ($p$Au, \sqsnn $=$ 38.7 \gev)~\cite{Schub:1995}, HERA-B ($p$C, $\sqsnn= 41.5 \gev$)~\cite{Abt:2008ya} and PHENIX ($pp$, $\sqs=200 \gev$)~\cite{PHENIXpp} experiments. The differential yields of \jpsi\ as functions of $y^*$ and $\pt$, obtained from \pAr\ data, are also shown in Fig.~\ref{fig::sig_jpsi}. Since the luminosity measurement is not available, only differential distributions with arbitrary normalization are shown. 

\begin{figure}
\begin{center}
\includegraphics[width=0.49\textwidth]{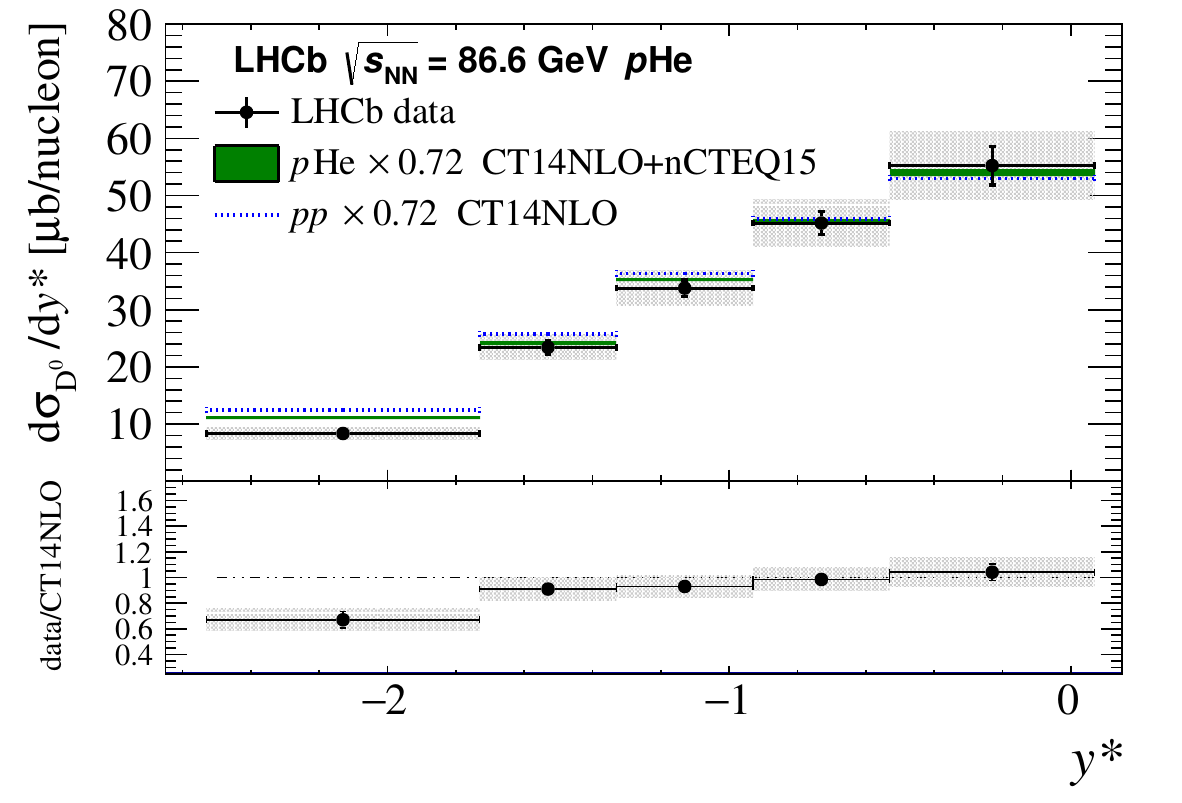}
\includegraphics[width=0.49\textwidth]{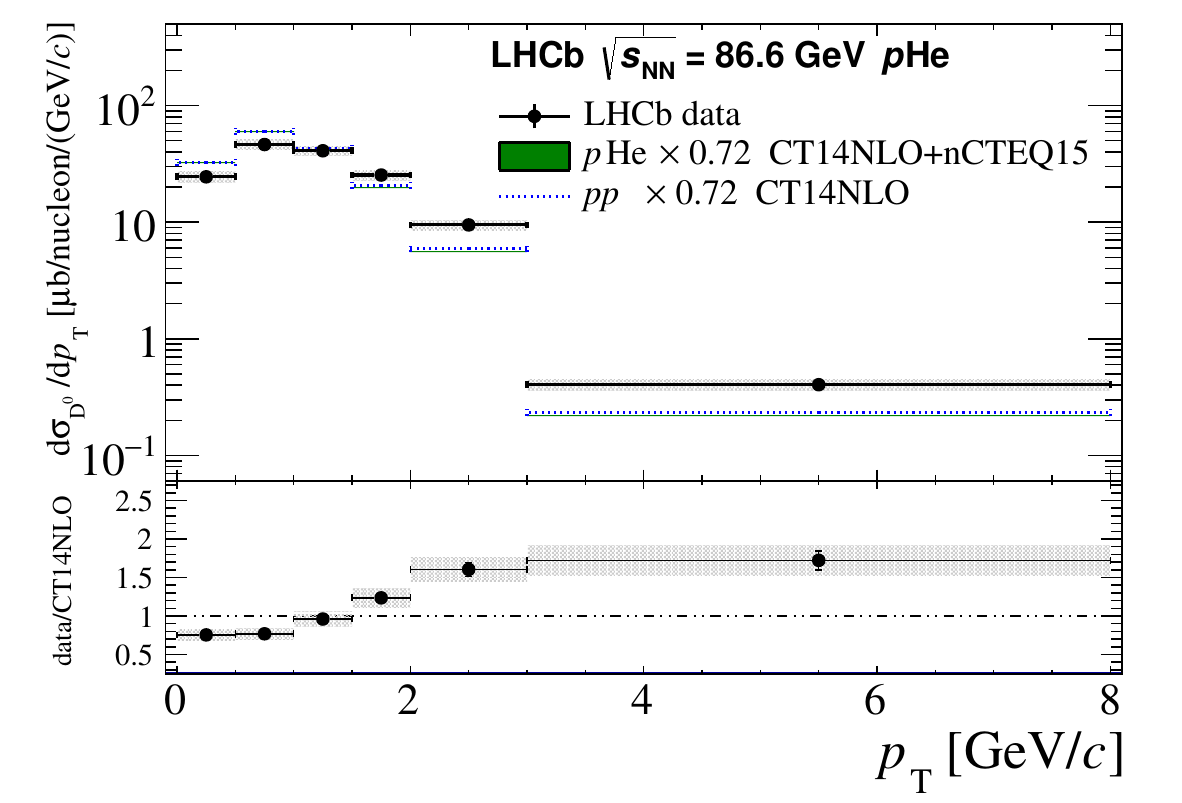}
\includegraphics[width=0.49\textwidth]{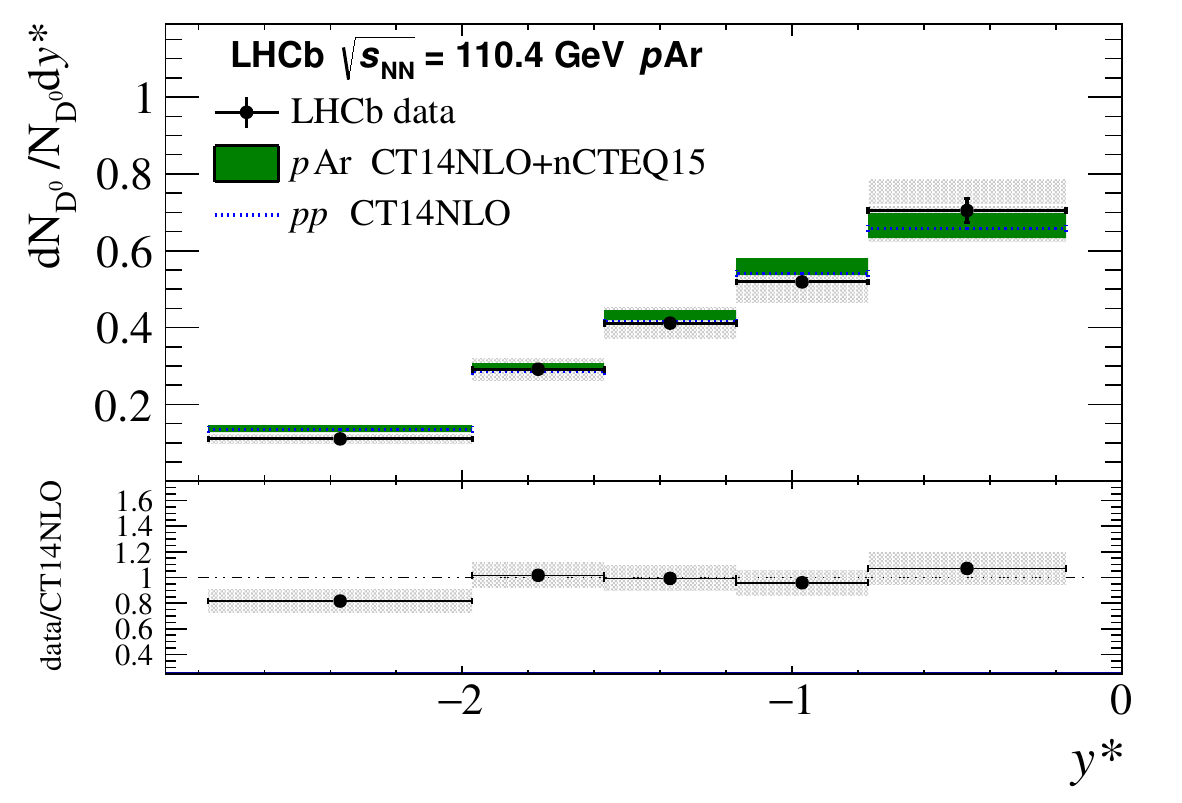}
\includegraphics[width=0.49\textwidth]{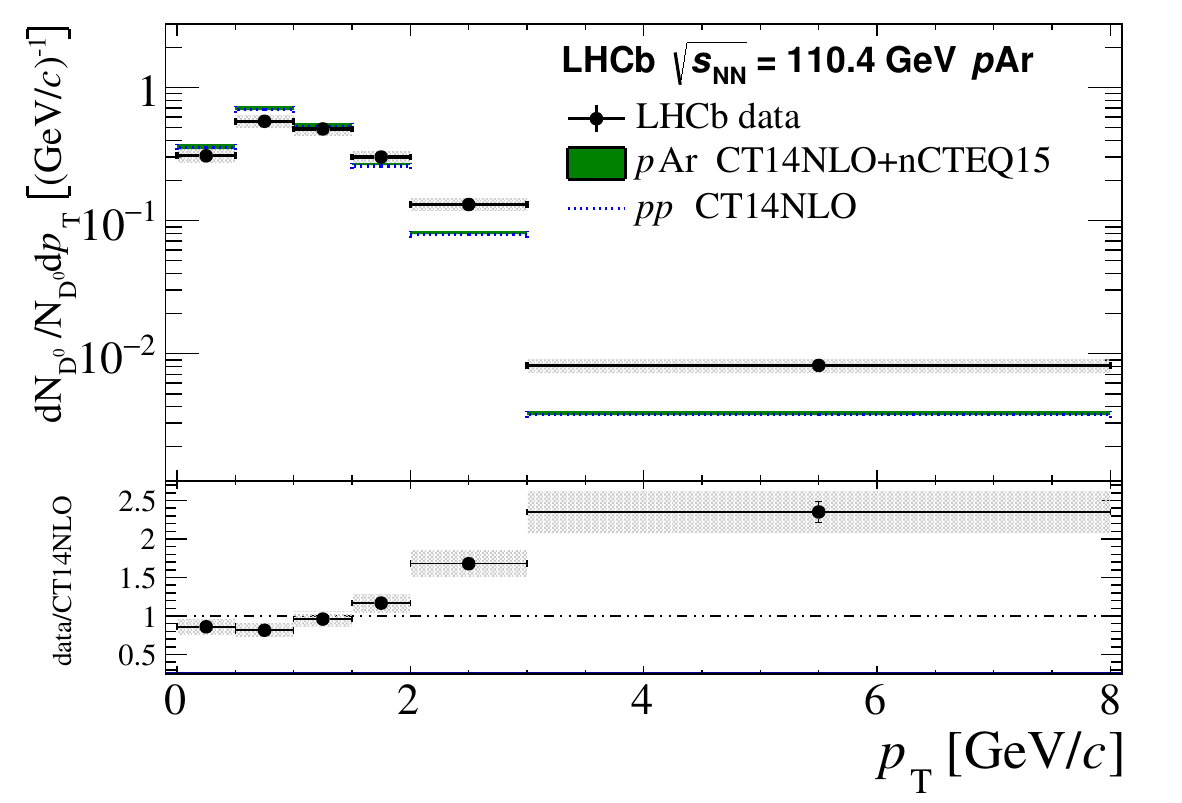}
\caption{Differential \Dz\ production cross sections for (top) \pHe\ and differential \Dz\ yields for (bottom) \pAr\ collisions, as a function of (left) center-of-mass rapidity $y^*$ and (right) transverse momentum $\pt$. The data points mark the bin centers. The quadratic sum  of statistical and uncorrelated systematic uncertainties are indicated by the vertical black lines. The correlated systematic uncertainties are indicated by the gray area. Theoretical predictions are described in the text. The lower panel of each plot shows the ratio of data to HELAC-ONIA $pp$ predictions.}
\label{fig::sig_D0}
\end{center}
\end{figure}

The \Dz\ differential cross sections per target nucleon obtained for the \pHe\ dataset, as functions of $y^*$ and $\pt$, are shown in Fig.~\ref{fig::sig_D0} and given in Ref.\cite{supplemental}. The HELAC-ONIA predictions overestimate the measured total cross section. The HELAC-ONIA predictions are rescaled by a factor 0.72 in Fig.~\ref{fig::sig_D0} to compare the shape of the distributions. Differential yields, with arbitrary normalization, of \Dz\ as functions of $y^*$ and $\pt$ obtained from \pAr\ data are also shown. 

In fixed-target configuration, the LHCb acceptance gives access to the large Bjorken-$x$ region of the target nucleon (up to $x\sim 0.37$ for \Dz mesons). 
In this region, because of the small number of nucleons in the helium nucleus, nuclear effects affecting \ccbar pairs are expected to be small.
On the other hand, as suggested in Refs.~\cite{Pumplin07, Dulat14}, the intrinsic charm contribution, based on a valencelike parton distribution, can be substantial at large Bjorken-$x$. Using the approximation for $x$, the fraction of the nucleon momentum carried by the target parton, 
\begin{equation}
x \simeq \frac{2 m_c}{\sqsnn}\exp(-y^*),
\end{equation}
where $m_c =1.28\gevcc$ is the mass of the $c$ quark\cite{cquarkmass}, 
the Bjorken-$x$ range $x\in [0.17,0.37]$ is obtained for the most backward bin. In this range any substantial intrinsic charm contribution should be seen in the \pHe\ results. As shown in Figs.~\ref{fig::sig_jpsi} and \ref{fig::sig_D0}, no strong differences are observed between \pHe\ data and the theoretical predictions which do not include any intrinsic charm contribution. Therefore, within uncertainties, no evidence of substantial intrinsic charm content of the nucleon is observed in the data. Future measurements with larger samples and more accurate theoretical predictions will permit to perform more quantitative studies, including the double-differential $[y^*, \pt]$ production cross section.  

In summary, we report the first measurement of heavy flavor production in fixed-target configuration at the LHC. The \jpsi\ and \Dz\ (including charge conjugate) production cross sections, measured in \pHe\ collisions at $\sqsnn=86.6 \gev$ in the rapidity range $[2,4.6]$, are found to be: \mbox{$\sigma_{\jpsi} = 652 \pm 33\stat \pm 42 \syst\nb/$nucleon} and \mbox{$\sigma_{D^0} = 80.8 \pm 2.4 \stat\pm 6.3 \syst\mub/$nucleon}. No evidence for a substantial intrinsic charm content of the nucleon is found.

\section*{Acknowledgements}
%
%
\noindent We express our gratitude to our colleagues in the CERN
accelerator departments for the excellent performance of the LHC. We
thank the technical and administrative staff at the LHCb
institutes.
We acknowledge support from CERN and from the national agencies:
CAPES, CNPq, FAPERJ and FINEP (Brazil); 
MOST and NSFC (China); 
CNRS/IN2P3 (France); 
BMBF, DFG and MPG (Germany); 
INFN (Italy); 
NWO (Netherlands); 
MNiSW and NCN (Poland); 
MEN/IFA (Romania); 
MSHE (Russia); 
MinECo (Spain); 
SNSF and SER (Switzerland); 
NASU (Ukraine); 
STFC (United Kingdom); 
NSF (USA).
We acknowledge the computing resources that are provided by CERN, IN2P3
(France), KIT and DESY (Germany), INFN (Italy), SURF (Netherlands),
PIC (Spain), GridPP (United Kingdom), RRCKI and Yandex
LLC (Russia), CSCS (Switzerland), IFIN-HH (Romania), CBPF (Brazil),
PL-GRID (Poland) and OSC (USA).
We are indebted to the communities behind the multiple open-source
software packages on which we depend.
Individual groups or members have received support from
AvH Foundation (Germany);
EPLANET, Marie Sk\l{}odowska-Curie Actions and ERC (European Union);
ANR, Labex P2IO and OCEVU, and R\'{e}gion Auvergne-Rh\^{o}ne-Alpes (France);
Key Research Program of Frontier Sciences of CAS, CAS PIFI, and the Thousand Talents Program (China);
RFBR, RSF and Yandex LLC (Russia);
GVA, XuntaGal and GENCAT (Spain);
the Royal Society
and the Leverhulme Trust (United Kingdom);
Laboratory Directed Research and Development program of LANL (USA).

\addcontentsline{toc}{section}{References}
\ifx\mcitethebibliography\mciteundefinedmacro
\PackageError{LHCb.bst}{mciteplus.sty has not been loaded}
{This bibstyle requires the use of the mciteplus package.}\fi
\providecommand{\href}[2]{#2}

\newpage

\centerline{\large\bf LHCb collaboration}
\begin{flushleft}
\small
R.~Aaij$^{27}$,
B.~Adeva$^{41}$,
M.~Adinolfi$^{48}$,
C.A.~Aidala$^{73}$,
Z.~Ajaltouni$^{5}$,
S.~Akar$^{59}$,
P.~Albicocco$^{18}$,
J.~Albrecht$^{10}$,
F.~Alessio$^{42}$,
M.~Alexander$^{53}$
% ...

{\footnotesize \it
% ...
$ ^{5}$Clermont Universit{\'e}, Universit{\'e} Blaise Pascal, CNRS/IN2P3, LPC, Clermont-Ferrand, France\\
% ...
$ ^{10}$Fakult{\"a}t Physik, Technische Universit{\"a}t Dortmund, Dortmund, Germany\\
% ...
$ ^{18}$INFN Laboratori Nazionali di Frascati, Frascati, Italy\\
% ...
$ ^{27}$Nikhef National Institute for Subatomic Physics, Amsterdam, Netherlands\\
% ...
$ ^{41}$Instituto Galego de F{\'\i}sica de Altas Enerx{\'\i}as (IGFAE), Universidade de Santiago de Compostela, Santiago de Compostela, Spain\\
$ ^{42}$European Organization for Nuclear Research (CERN), Geneva, Switzerland\\
% ...
$ ^{48}$H.H. Wills Physics Laboratory, University of Bristol, Bristol, United Kingdom\\
% ...
$ ^{53}$School of Physics and Astronomy, University of Glasgow, Glasgow, United Kingdom\\
% ...
$ ^{59}$University of Cincinnati, Cincinnati, OH, United States\\
% ...
$ ^{61}$Syracuse University, Syracuse, NY, United States\\
% ...
$ ^{73}$University of Michigan, Ann Arbor, United States, associated to $^{61}$\\
% ...
}
\end{flushleft}
\end{document}